\documentclass[10pt,letterpaper]{article}
\usepackage[top=0.85in,left=2.75in,footskip=0.75in]{geometry}

\usepackage{amsmath,amssymb,amsfonts}
\usepackage{bm}
\usepackage{bbm}

\usepackage{graphicx}

\usepackage{algorithm,algorithmic}

\usepackage{changepage}

\usepackage{textcomp,marvosym}

\usepackage{cite}

\usepackage{nameref,hyperref}

\usepackage[nopatch=eqnum]{microtype}
\DisableLigatures[f]{encoding = *, family = * }

\usepackage[table]{xcolor}

\usepackage{array}

\newcolumntype{+}{!{\vrule width 2pt}}

\newlength\savedwidth

\newcommand\thickhline{\noalign{\global\savedwidth\arrayrulewidth\global\arrayrulewidth 2pt}%
\hline
\noalign{\global\arrayrulewidth\savedwidth}}

\usepackage{setspace} 
\raggedright
\usepackage[aboveskip=1pt,labelfont=bf,labelsep=period,justification=raggedright,singlelinecheck=off]{caption}

\makeatletter
\renewcommand{\@biblabel}[1]{\quad#1.}
\makeatother

\usepackage{lastpage,fancyhdr,graphicx}
\usepackage{epstopdf}
\fancyheadoffset[L]{2.25in}
\fancyfootoffset[L]{2.25in}
\begin{document}
\vspace*{0.2in}

\begin{flushleft}
{\Large
\textbf\newline{Improving opioid use disorder risk modelling through behavioral and genetic feature integration} 
}
\newline
\\
Sybille L\'egitime\textsuperscript{1*},
Kaustubh Prabhu\textsuperscript{1},
Devin McConnell\textsuperscript{1},
Bing Wang\textsuperscript{1},
Dipak K. Dey\textsuperscript{2},
Derek Aguiar\textsuperscript{1,3*}
\\
\bigskip
\textbf{1} School of Computing, University of Connecticut, Storrs, CT, United States
\\
\textbf{2} Department of Statistics, University of Connecticut, Storrs, CT, United States
\\
\textbf{3} Institute for Systems Genomics, University of Connecticut, Storrs, CT, United States
\\
\bigskip

%
%





* sybille.legitime@uconn.edu
* derek.aguiar@uconn.edu

\end{flushleft}

\section*{Abstract}
Opioids are an effective analgesic for acute and chronic pain, but also carry a considerable risk of addiction leading to millions of opioid use disorder (OUD) cases and tens of thousands of premature deaths in the United States yearly.
Estimating OUD risk prior to prescription could improve the efficacy of treatment regimens,  monitoring programs, and intervention strategies, but risk estimation is typically based on self-reported data or questionnaires.
We develop an experimental design and computational methods that combine genetic variants associated with OUD with behavioral features extracted from GPS and Wi-Fi spatiotemporal coordinates to assess OUD risk.
Since both OUD mobility and genetic data do not exist for the same cohort, we develop algorithms to (1) generate mobility features from empirical distributions and (2) synthesize mobility and genetic samples assuming an expected level of disease co-occurrence.
We show that integrating genetic and mobility modalities improves risk modelling using classification accuracy, area under the precision-recall and receiver operator characteristic curves, and $F_1$ score.
Interpreting the fitted models suggests that mobility features have more influence on OUD risk, although the genetic contribution was significant, particularly in linear models.
While there exist concerns with respect to privacy, security, bias, and generalizability that must be evaluated in clinical trials before being implemented in practice, our framework provides preliminary evidence that behavioral and genetic features may improve OUD risk estimation to assist with personalized clinical decision-making.



\section*{Introduction}
Opioids are a class of drugs that target opioid receptors to treat chronic and acute pain. 
While opioids can be successfully administered without substantial adverse effects, they have caused significant negative impacts on many individuals and society, more generally. 
In 2020, an estimated 2.7 million people suffered from opioid use disorder (OUD)~\cite{nih2021}, which is a clinical diagnosis that combines opioid abuse (intermittent and reversible use) with physical dependence and tolerance-building accompanied by adverse side-effects or addiction~\cite{yu1997development,hasin2013dsm}. 
In 2021, 80,411 people died from overdose involving opioids, with nearly 21\% of those deaths involving prescription opioids~\cite{wonder2022}. 
That same year, the prevalence of problematic opioid use in patients with chronic non-cancer pain was estimated to be as high as 36.3\%~\cite{jantarada2021prevalence}.
While patient outcomes are highly heterogeneous in response to opioid therapy, the medical treatment of physical pain or emotional stress are common pathways that lead to OUD development~\cite{stumbo2017patient}.
Thus, the risk of developing OUD is an important factor to consider before determining personalized treatment strategies, which may include prioritizing alternative analgesics or designing early intervention mechanisms.

Disease risk varies both between individuals in a population and per individual over time due to inherent~\cite{li2022atherospectrum} and dynamically changing risk factors, respectively~\cite{giannoula2018identifying}.
Previous studies have identified factors to estimate both inherent and dynamic OUD risk, including structured interviews and risk assessment questionnaires like the Screener and Opioid Assessment for Patients with Pain (SOAPP)~\cite{butler_validation_2004,martel_catastrophic_2013}, the revised SOAPP~\cite{butler_validation_2008}, and the short form SOAPP~\cite{koyyalagunta_risk_2013}; though prior work suggests that these existing OUD risk factors are insufficient to predict individual OUD risk~\cite{turk_predicting_2008}. 
More recently, machine learning methods using administrative claims data, like prescription use, sociodemographics, and health status, have achieved high negative predictive value ($99.9\%$), but low precision ($0.18\%$) for identifying opioid overdose risk~\cite{lo-ciganic_evaluation_2019}.

The interindividual variability in OUD prevalence has been characterized through a more personalized approach in disease association and heritability studies. 
Genome-wide association studies (GWAS) have identified many single nucleotide polymorphisms (SNPs) and copy number variations associated with opioid abuse and addiction~\cite{cheng2018genome,nelson_evidence_2016,gelernter_genome-wide_2014,li_genome-wide_2015}.  
Inherent OUD risk can be computed by a weighted sum of genetic risk alleles that were identified by prior genome-wide association studies on the phenotype of interest (i.e., a polygenic risk score -- PRS)~\cite{iyegbe2014emerging,choi2020tutorial}.
While PRS have clinical utility for heritable diseases~\cite{mavaddat2019polygenic}, particularly for prevention~\cite{lambert2019towards}, diseases that have significant liability due to environmental factors present a notable challenge for PRS~\cite{iyegbe2014emerging}.  
Non-genetic static factors, such as sex or population, are routinely incorporated into PRS, but have limited utility when considering the high association of substance use disorders with behavioral patterns~\cite{witkiewitz2022mechanisms}.
Moreover, disease risk depends on social and familial features that change over time and may influence disease risk directly or through gene-by-environment interactions~\cite{whitesell2013familial}.

A recent perspective on evaluating the environmental component of risk, disease progression, and treatment outcomes uses \textit{mobility features} derived from wearable devices and GPS and Wi-Fi traces~\cite{canzian_trajectories_2015,yue_fusing_2018}. 
These mobility features, like entropy of movement or total distance traveled, are informative of behavior, and understanding how disease risk changes in response to variability in activity can inform treatment decisions (pre-diagnosis), intervention strategies (during treatment), and recovery monitoring (post-treatment). 
Predictive models trained on smartphone GPS, and Wi-Fi~\cite{farhan_behavior_2016,yue_fusing_2018} and cell phone usage data~\cite{saeb_mobile_2015} have been used to predict clinical depression as evaluated through Patient Health Questionnaires (PHQ) and clinical assessments.
Mobility has also been used to detect depressive and manic episodes~\cite{gruenerbl_using_2014}, classify behavior and mental health outcomes~\cite{wang_studentlife:_2014,ben-zeev_next-generation_2015}, and monitor mental health for patients with schizophrenia~\cite{wang_patterns_2016}. 
Importantly, mobility features have been highly effective in evaluating disease susceptibility or periods of high disease severity for conditions that frequently co-occur with OUD~\cite{palmius_detecting_2017,canzian_trajectories_2015}.
However, these approaches only consider the environmental component of disease risk while ignoring  genetic risk factors, which are an important source of phenotypic variance in substance use disorders~\cite{ducci2012genetic}.

\begin{figure}[!h]
\centering
\includegraphics[width=1.0\textwidth]{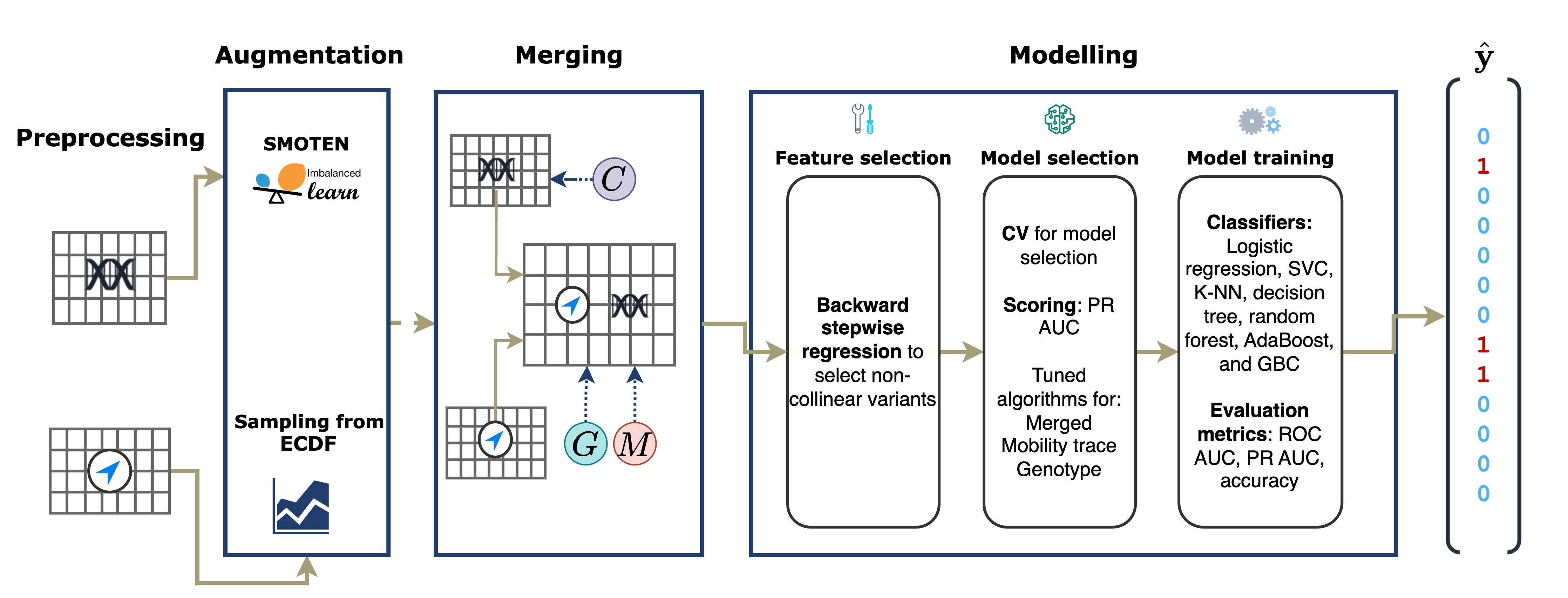}
\caption{{\bf Overview of our integrative approach for estimating disease risk.}
The mobility trace and genetic data are \textit{preprocessed}, then \textit{augmented} to balance the genetic and mobility trace sample sizes. 
     The augmented data is \textit{merged} using a disease co-occurrence parameter ($C$), genetic relative risk ($G$), and mobility relative risk ($M$). 
     In the \textit{modelling} step, features and models are selected, and classifiers are trained to estimate OUD risk.}
\label{fig:overview}
\end{figure}

In this work, we present the first approach that combines genetic and mobility features to estimate disease risk.
Leveraging how clinical biomarkers, environmental factors, and genetic features interact to affect substance use disorders~\cite{barr2022clinical,kinreich2021predicting}, our framework for risk prediction combines genetic and environmental modalities with a focus on mobility features that are extracted from mobility traces. 
Since there is currently no available dataset that combines mobility traces with genetic data, we develop (a) data augmentation algorithms to synthesize mobility trace samples from mobility feature empirical distributions and (b) data fusion algorithms based on assumptions about disease penetrance and expected disease co-occurrence~(Fig.~\ref{fig:overview}).
After combining genetic and mobility features, we train a suite of classifiers to estimate OUD risk in a variety of experimental scenarios and show that combining genetic and mobility modalities improves risk modelling using classification accuracy, area under the precision-recall and receiver operator characteristic curves (AUPRC, AUROC), and $F_1$ score.
Extensive interpretations of fitted models suggests that mobility features have more influence on disease risk, although the genetic contribution was significant, particularly in linear models.
We conclude with a comprehensive discussion of implementation concerns with respect to privacy, security, bias, explainable AI, and generalization to other diseases and patient scenarios. 
Our source code and scripts are freely available at \href{https://github.com/bayesomicslab/OUD-Risk-Prediction}{https://github.com/bayesomicslab/OUD-Risk-Prediction}.


\section*{Materials and methods}

Our approach begins with extracting mobility features from GPS and Wi-Fi spatiotemporal coordinates, and genetic features from genome-wide association study data~(Fig.~\ref{fig:overview}).
These mobility features include $9$ novel features based on the location type extracted from Google Places.
In the \textit{pre-processing} and \textit{augmentation} phases, we perform standard quality control measures on both the genetic and mobility data, extract features, leverage an existing data augmentation method to synthesize genetic data, and generate mobility features using a new simulation algorithm. 
Since mobility data do not exist for OUD populations, the subsequent \textit{merging} phase creates hybrid (combined) observations by sampling a case or control genetic sample at random, appending features of a sample case or control mobility trace based on an assumed co-occurrence probability, and assigning case status to the hybrid samples using a disease model based on relative risk. 
The resulting samples from the pre-processing, augmentation, and merging phases enable the creation of datasets with three distinct feature sets: genetic data only, mobility traces only, and combined.
In the final \textit{modelling} phase we (a) perform feature selection on the genetic data, (b) select model parameters using cross-validation (CV), and train a diverse set of classifiers to estimate OUD risk.


\subsubsection*{Genetic features}

Variability in the genome is characterized by the sequences of alternative forms of genetic variants (alleles), or \textit{genotypes}, typically collected by genotyping arrays~\cite{dunning2007beadarray} or DNA sequencing~\cite{evrony2021applications}.
Let the genotype matrix $\bm{X}^g$ and label vector $\bm{y}^g$ be $\bm{X}^g=(x_{ij}^g)_{i=1,j=1}^{i=N_g,j=L_g}$ and $\bm{y}^g=(y_{i}^g)_{i=1}^{i=N_g}$, respectively, where $N_g$ is the number of genetic samples, $L_g$ is the number of genetic variants, $x_{ij} \in \{0,1,2\}$ denotes the count of the minor allele for sample $i$ and position $j$, and $y_{i}^g \in \{0,1\}$ denotes disease status, where 0 and 1 indicate control and case status, respectively.
These data can be agnostic to prior GWAS, in which case $L_g \approx 10^6$, or be defined based on the variants that have prior associations with the disease of interest ($L_g \approx 10^2$).

\subsubsection*{Mobility features}
Mobility traces are temporally sequenced latitude and longitude positions extracted from the participant phones using GPS data or Wi-Fi association records. 
Mobility features are computed from mobility trace data.
Let the mobility feature matrix and label vector be $\bm{X}^m=(x_{ij}^m)_{i=1,j=1}^{i=N_m,j=L_m}$ and $\bm{y}^m=(y_{i}^m)_{i=1}^{i=N_m}$, respectively, where $N_m$ is the number of mobility trace samples, $L_m$ is the number of mobility features, $x_{ij} \in \mathbb{R}$ denotes the $j^{th}$ mobility feature value for sample $i$ and $y_{i}^m \in\{0,1\}$ is the disease status, where 0 and 1 indicate control and case status, respectively.
We consider a total of $L_m=21$ features (Table~\ref{tab:labels}) that have demonstrated utility for diagnosing diseases that often co-occur with OUD~\cite{farhan_behavior_2016,yue_fusing_2018,saeb_mobile_2015,canzian_trajectories_2015,palmius_detecting_2017,rinzivillo2014purpose}.

Mobility features can be categorized as either movement-based or location-based.

\begin{table}[!ht]
\centering
\caption{\textbf{Legend of mobility feature abbreviations and names.}}
\label{tab:labels}
\begin{tabular}{|l|l|}
\hline
\textbf{Abbreviation} & \textbf{Name}\\ 
\thickhline
VAR &  Location variance\\
AVG SPD & Average moving speed (km/h) \\
ENT & Entropy \\
NORM ENT & Normalized entropy \\
HOME & Time spent at home\\
TRANS TIME & Transition time\\
TOT DIST & Total distance travelled\\
ROUT IND & Routine index\\
INDGR & Indegree\\
OUTDGR & Outdegree\\
UNIQUEC & Number of unique locations visted\\
UNITC & Unique cluster type\\
OUTDOORS\_REC & `Outdoors \& Recreation' cluster\\
PROFESS\_OTH & `Professional \& Other Places' cluster\\
SHOP & `Shop \& Service' cluster\\
FOOD & `Food' cluster\\
TRANSPORT & `Travel \& Transport' cluster\\
RESIDENCE & `Residence' cluster\\
UNIVERSITY & `College \& University' cluster\\
ARTS\_ENTERT & `Arts \& Entertainment' cluster\\
NIGHTLIFE & `Nightlife Spot' cluster\\ 
\hline
\end{tabular}
\end{table}

\paragraph{Movement-based features}
Movement-based features capture the activity of an individual based on their positional trajectories.
We consider $3$ movement-based features.
The \textit{average moving speed} in meters per second is the instantaneous speed estimated from adjacent positions. 
\textit{Transition time} refers to the proportion of time spent in transition by sample $i$ (i.e. with moving speed $>1$ km per hour).
We also compute the \textit{total distance travelled} as the sum of Harversine distances between adjacent positions.

\paragraph{Location-based features}
Location-based features represent how an individual interacts with discrete locations. 
First, distinct locations are identified by clustering the sample latitude and longitude positions.
Then, distinct location clusters are labeled using the Google Places API, which includes bounding boxes for discrete places and $9$ categories: outdoors and recreation, professional \& other places, shop \& service, food, travel \& transport, residence, college and university, arts \& entertainment, and nightlife spots.
We consider $18$ location-based features in total.

Given the empirical probability of sample $i$ being in location $k$ as $p_{ik}$, the \textit{entropy} of sample $i$ is defined as $-\sum_{k} p_{ik} \log p_{ik}$ and measures the uncertainty in the location of sample $i$.
Since entropy increases with the number of unique locations visited, we also compute the \textit{cluster normalized entropy},   
\begin{equation*}
\frac{-\sum_{k} p_{ik} \log p_{ik}}{\log N_i^{C}},
\end{equation*}
where $N_i^{C}$ is the number of clusters for sample $i$.
We label the location where the sample spent the most time between the hours of $12$am to $6$am as their home.
The \textit{time spent at home} measures the average amount of time spent at home per day.
\textit{Location variance} measures the positional variability of a sample as $\log(\sigma^2_{lon}+\sigma^2_{lat})$, where $\sigma^2_{lon}$ and $\sigma^2_{lat}$ are the empirical variance of the longitude and latitude, respectively.
We also consider the \textit{routine index}, which quantifies the regularity of places visited over time.
We compute the ratio of time spent at location $k$ in week $w$ to time spent at the same location in week $w+1$ for each location and week.
The routine index is the average of these ratios across all locations and times.
Additionally, we compute the average number of times a sample \textit{left} or \textit{entered} a location, the average \textit{time spent} in each location, and the \textit{number of unique locations} visited.
Finally, we compute $9$ features based on the average time spent in each of the Google Places API location categories.

\subsubsection*{Generating synthetic mobility feature samples}

\begin{figure}[!h]
\centering
\includegraphics[width=1.0\textwidth]{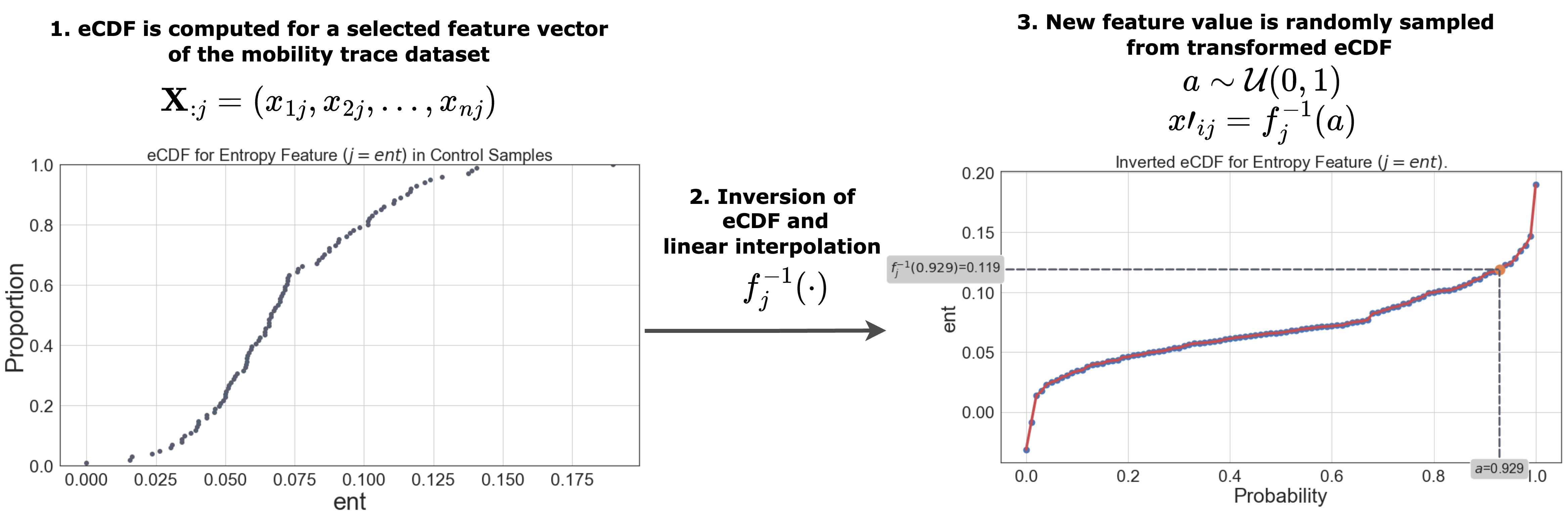}
\caption{{\bf Mobility trace features simulation.}
(1) From a pre-selected population (case or control), the eCDF for a feature is generated. (2) The eCDF for that feature is inverted, and linear interpolation is performed between points. (3) The simulated feature value is created by random sampling of the inverse eCDF. The process is repeated across all features and all samples in the population.}
\label{fig:sim}
\end{figure}

Due to the difficulties associated with collecting mobility traces, it is often the case that $N_m<<N_g$; therefore, combining mobility and genetic data requires generating new samples of mobility traces or features.
However, mobility features are typically modelled conditionally, not generatively, since they have unknown prior distributions and complex interdependencies. 
We address these challenges by developing a simulation algorithm that samples mobility features based on their smoothed empirical cumulative distribution functions (eCDF)~(Fig.~\ref{fig:sim}). 
We separate case mobility trace samples from controls, compute the eCDF $f_j(\cdot)$ of each mobility feature independently, and linearly interpolate between each observation.
We generate a new mobility sample $i$ by drawing $a \sim \mathcal{U}(0,1)$ and then sampling new mobility feature $x'_{ij}=f_j^{-1}(a)$ for $j=1,\dots,L_m$.

\subsubsection*{Hybrid sample simulator and risk models}

Since there exists no mobility data for individuals with OUD, we generate hybrid samples -- a combination of genetic and mobility features -- by considering the co-occurrence factor $C$ between the diseases in each dataset. 
We choose clinical depression as the co-occurring disease, since the prevalence of psychiatric disorders in individuals treated for OUD suggests a non-negligible association between the two conditions~\cite{ZHU2021108996, Schepis};  with observed co-occurrences for OUD and depression of 26.54\% in mental health patients~\cite{vekaria_association_2021}, and 3.4\% to 81\% among chronic pain patients~\cite{VANRIJSWIJK201937}.
Let $C$ define the expected co-occurrence probability of a disease with an indexed disease.
Clearly, if both datasets consider the same disease, then $C=1$, but if the diseases are related yet distinct (e.g., OUD and depression) $C \leq 1$. 
Here, OUD is the indexed disease, and depression represents the co-occurring disorder.

Additionally, we use relative risk (RR) to model variability in disease penetrance.
The RR is the probability of the outcome for one exposure group divided by the probability of the outcome of a reference exposure group \cite{mcnutt_estimating_2003}. 
For example, an RR of $10$ indicates that the exposure presents 10 times the risk of the outcome than the reference group.
Parameterizing co-occurrence and RR enables our simulator to generate a higher diversity of realistic scenarios where both the dependence between diseases and the proportion of variance explained by genetic and environmental components can be adjusted. 

Our simulator requires specification of a co-occurrence $C$, genotype RR $G$, mobility trace RR $M$, and a number of samples per case and control group $n$~(Algorithms~\ref{alg:cap} and \ref{alg:cap2}).
While the number of samples per case and control group is less than $n$, we sample or generate a genotype vector $\hat{\bm{x}}^g_i$ uniformly at random across case and control groups.
Next, we sample or generate a mobility feature vector $\hat{\bm{x}}^m_i$ from the same or different disease group according to the co-occurrence $C$.
Let function $\mathbbm{1}(\hat{\bm{x}}_i)$ evaluate to $1$ if sample $\hat{\bm{x}}_i$ is a case and $0$ otherwise. 
We draw two random variables $a^g \sim \text{Bern}(G/(G+1))$ and $a^m \sim \text{Bern}(M/(M+1))$.
We concatenate $\hat{\bm{x}}^g_i$ and $\hat{\bm{x}}^m_i$ to produce $\hat{\bm{x}}_i$ and set its disease status to $\mathbbm{1}(\hat{\bm{x}}^g_i)\cdot a^g \lor \mathbbm{1}(\hat{\bm{x}}^m_i)\cdot a^m$.
Given $\hat{\bm{X}}=(\hat{\bm{x}}_i)_{i=1}^n$, we consider $7$ models to predict OUD status: logistic regression (LOGIT), linear SVC (SVC), K-nearest neighbors (KNN), decision tree (DT), random forest (RF), AdaBoost (ADA), and gradient boosting classifiers (GBC)~\cite{scikit-learn}.

\begin{algorithm}[h!]
\small
\caption{Algorithm for generating hybrid dataset}
\label{alg:cap}
\begin{algorithmic}
\STATE \textbf{Inputs}: Number of hybrid samples with a \textit{case} status (n\_cases), Number of hybrid samples with a \textit{control} status (n\_controls), target number of case and control samples ($n$), genotype data and labels ($\bm{X}^g, \bm{y}^g$), mobility trace data and labels ($\bm{X}^m, \bm{y}^m$), co-occurrence $C$, mobility trace relative risk $M$, genotype relative risk $G$
\STATE \textbf{Output}: hybrid\_data ($\hat{\bm{X}}, \hat{\bm{y}}$)
\WHILE{n\_cases or n\_controls $\leq$ $n$}
\STATE $\hat{\bm{x}}^g_i \sim s(\mathcal{U}(0,1))$ \COMMENT{Select case or control genotype at random}
\STATE $\hat{\bm{x}}^m_i \gets \text{simulate\_mt\_sample}^\ast$ according to $C$ \COMMENT{see Alg. \ref{alg:cap2}}
\STATE $\mathbbm{1}(\hat{\bm{x}}_i) \gets 1$ if $\bm{x}^g_i$ is a case $0$ otherwise
\STATE $a^g \sim \text{Bern}(G/(G+1))$ 
\STATE $a^m \sim \text{Bern}(M/(M+1))$
\STATE hybrid\_sample $\gets$ concatenate($\hat{\bm{x}}^g_i, \hat{\bm{x}}^m_i$)
\STATE $\hat{y} \gets$ $\mathbbm{1}(\hat{\bm{x}}^g_i)\cdot a^g \lor \mathbbm{1}(\hat{\bm{x}}^m_i)\cdot a^m$
\STATE hyrbid\_data.push(hybrid\_sample, $\hat{y}$)
\ENDWHILE
\end{algorithmic}
\end{algorithm}

\begin{algorithm}[H]
\small
\caption{$\text{simulate\_mt\_sample}^\ast$}
\label{alg:cap2}
\begin{algorithmic}
\STATE \textbf{Input}: $\boldsymbol{X}^m$ \COMMENT{Case or control mobility samples}
\STATE \textbf{Output}: synthetic\_sample\_output \COMMENT{Data point created by simulator}
\STATE $N^m \gets$ total number of mobility trace samples
\STATE $L^m \gets$ total number of mobility trace features
\FOR{$j$ in $L^m$}
\STATE $\text{feature\_vector} \gets \bm{v}_{N^m, j}^m$ 
\STATE $f^{-1}(\cdot) \gets$  Inverted and interpolated eCDF of feature\_vector
\STATE $a \sim \mathcal{U}(0,1)$ \COMMENT{Random sample from uniform distribution}
\STATE synthetic\_sample\_output.append($f^{-1}(a)$) \COMMENT{Add feature to synthetic sample}
\ENDFOR
\end{algorithmic}
\end{algorithm}

\section*{Results}

For the genetic samples, we used the substance use dependence Only (SUD) and substance use dependence, childhood trauma and related disorders (SUDCT) consent groups from the Genome-Wide Association Study of Heroin Dependence ($N_g=2883$; study accession \href{https://www.ncbi.nlm.nih.gov/projects/gap/cgi-bin/study.cgi?study_id=phs000277.v1.p1}{phs000277.v1.p1}).
The case samples met the Diagnostic and Statistical Manual of Mental Disorders (DSM-IV) criteria for heroin dependence (a type of opioid), and controls were individuals assessed as not meeting the  DSM-IV criteria for heroin dependence along with unassessed population controls. 
We computed mobility features from the LifeRhythm study, which  focused on the prediction of clinical depression~\cite{farhan_behavior_2016,yue_fusing_2018}; the study recorded location data every minute from University of Connecticut student participant phones over $8$ months (October 2015 to May 2016) using the LifeRhythm app. 
LifeRhythm ran as a background process, collecting three types of sensing data: GPS location data, motion-processor-generated activity data, and Wi-Fi association records that indicate the association status of a smartphone with a wireless access point.
LifeRhythm study samples were mapped to randomly generated identifiers and the mapping table between identifiers and real sample identities was deleted to preserve sample anonymity. 
Mobility features were extracted from the latitude and longitude locations of the sensing data and aggregated for each unique sample identifier, yielding $N_m=144$ samples.
Informed consent was secured from participants in both prior studies and approval from the University of Connecticut Institutional Review Board was obtained prior to data access (reference number H20-0018)~\cite{farhan_behavior_2016,maloney2007suicidal}.

\subsection*{Data processing}

\subsubsection*{Genetic data processing}

We processed the genetic data using PLINK~\cite{choi_tutorial_2020} by removing duplicate features and variants that (a) had minor allele frequency less than $0.01$, (b) violated Hardy-Weinberg Equilibrium ($\chi^2$ test; p-value less than $10^{-7}$), or (c) were missing in more than 1\% of the samples. Further, we removed samples that contained more than $1\%$ missing variants, where reported sex did not match inferred sex, or that identified as having a relative in the data. The original sample size of $N_g=2883$ was reduced to in $N_g=2671$ as a result of this quality control step.
We imputed missing genomic variants by their mode and we identified SNPs that presented significant associations with OUD (EFO$\_$0010702) and opioid dependence (EFO$\_$0005611) from the NHGRI-EBI GWAS Catalog~\cite{hastings_burdett_2017}.
Because most of the GWAS catalog reference variants could not be extracted from the heroin dependence genotype samples, \textit{proxy} variants were identified as SNPs showcasing high linkage disequilibrium with catalog variants from all African and European subpopulations in the 1000 Genomes project, phase 3 data (D\_prime=1.0, window size=300kb, and $R^2=0.8$)~\cite{ensembl2021}. The resulting genotype matrix $\bm{X}^g$ --extracted using PLINK-- included only the proxy variants ($L_g=33$).

\subsubsection*{Mobility trace data processing}
For the mobility data, we identified discrete locations visited by an individual by running the density-based clustering algorithm DBSCAN. Through grid search CV, we selected DBSCAN hyperparameters $\epsilon$, the minimum number of points to define a cluster, the distance metric, and the algorithm to compute the nearest neighbors. With the objective of maximizing the \textit{silhouette score}, we set the model hyperparameter ranges to be $ 0.002 \leq \epsilon \leq 0.056$ and $3 \leq$ \verb|min_pts| $\leq 161$ in increments of $0.002$ and $2$, respectively.
The selected hyperparameters were $\epsilon = 0.054$, \verb|min_pts| $=3$, the \verb|haversine| metric, and \verb|ball_tree| algorithm; $\epsilon$ and the sample coordinates were converted to radians to comply with scikit-learn's \verb|haversine| metric requirement of radian units.
Distinct clusters were labelled using the Google Places API, which includes bounding boxes for discrete places and $9$ categories: outdoors and recreation, professional \& other places, shop \& service, food, travel \& transport, residence, college and university, arts \& entertainment, and nightlife spots.
If a cluster centroid was not within the bounding box of any known place, we matched it with the bounding box closest to the cluster centroid using a K-D tree implementation. 
We then generated the mobility feature matrix $\bm{X}^g$ by computing the aforementioned mobility features~\cite{martin1996density} ($L^m=21$).


\subsubsection*{Hybrid data processing}

Since we observed imbalance in the case-control class across consent groups in the genetic data ($n_{cases} = 2170$, $n_{controls} = 501$), we oversampled controls using the synthetic minority oversampling technique for categorical variables (SMOTEN)~\cite{Chawla_2002}.
We generated $100$ random datasets using our hybrid sample simulator with number of samples $N=2000$ ($1000$ cases and controls).
We varied the co-occurrence probabilities $C \in \{0.6,0.8,1.0\}$, genetic and mobility RR $(G,M) \in \{(\infty,\infty),(15,5),(15,1),(10,5),(10,1),(5,5)\}$, and feature sets, including genetic data only, mobility traces only, and combined features. 
We standardized features by subtracting the empirical means and dividing by the empirical standard deviations of the training data.

\subsection*{Model selection and evaluation}

We selected genetic features for each expected co-occurrence level and RR configuration in held-out datasets to remove collinear features.
Hyperparameters were selected per disease risk score model and feature set using 10-fold CV and average precision.
We evaluated each model on the $100$ datasets for each feature set, expected co-occurrence, and RR configuration using AUROC, AUPRC, classification accuracy, and $F_1$ score assuming a $0.5$ probability threshold. 
We interpreted fitted models using SHapely Additive exPlanations (SHAP), which estimates the change in the expected model prediction when conditioning on feature $j$ for a specific sample~\cite{NIPS2017_7062}.

\subsection*{Risk score model analysis}

First, we evaluated the risk score models across the three feature sets with selected co-occurrence $C=0.8$, and complete disease penetrance, $(G,M)=(\infty,\infty)$.
In simpler methods that did not include interaction effects (e.g., LOGIT), the models trained on merged data outperformed models trained on either modality (Table~\ref{tab:rocauc}). 
In more complicated models (e.g., boosting classifiers), models trained on the merged data were indistinguishable from the same model trained on mobility features, suggesting that mobility traces are sufficient to rank samples based on disease state probabilities.
The ensemble methods had significantly higher AUROC across the three feature sets: GBC in merged and mobility feature sets (Welch's two sample t-test; p-values $\leq 2.93  \times 10^{-8}$) and random forests in the genotype set (Welch's two sample t-test; p-values $\leq 2.36 \times 10^{-6}$).
Additionally, AUPRC and $F_1$ scores showed similar patterns with respect to the performance of the ensemble methods (Tables~\ref{tab:prauc} and \ref{tab:f1_inf}). 
Models that made linear and additive effects assumptions performed markedly worse than the models that allow for more complex interactions between features, indicating that interactions between genetic variants, mobility features, or gene-by-environment may be present; this interpretation is consistent with the literature that suggests there exists significant gene-by-environment interactions in substance use disorders~\cite{vink2016genetics}.

\begin{table}[!ht]
\centering
\caption{AUROC scores across models and feature sets.}
\setlength{\tabcolsep}{5pt}
\begin{tabular}{|l+l|l|l|l|}
\hline
& \multicolumn{3}{c|}{\textbf{Dataset}} \\ \cline{2-4}
\textbf{Model} & \textbf{Merged} & \textbf{Mobility} & \textbf{Genotype} \\ 
\hline
LOGIT & 0.68 (2.5) & 0.63 (2.7) & 0.63 (2.7)\\
SVC & 0.68 (2.5) & 0.63 (2.7) & 0.63 (2.7)\\
KNN & 0.69 (2.6) & 0.71 (2.6) & 0.66 (3.2)\\
DT & 0.75 (3.1) & 0.75 (3.6) & 0.64 (2.5)\\
RF & 0.88 (1.7) & 0.88 (1.8) & \textbf{0.68 (2.9)}\\
ADA & 0.89 (1.8) & 0.89 (1.9) & 0.63 (2.7)\\
GBC & \textbf{0.91 (1.5)} & \textbf{0.91 (1.8)} & 0.67 (2.8)\\ \hline
\end{tabular}
\begin{flushleft} Mean AUROC and standard errors ($\times 10^{-2}$) across 100 datasets with co-occurrence $C=0.8$ and relative risks $(G,M)=(\infty,\infty)$.
Bold values indicate models with significantly better AUROC than other methods within a feature set (Welch's two sample t-test; all p-values $\leq 2.36 \times 10^{-6}$).
\end{flushleft}
\label{tab:rocauc}
\end{table}

\begin{table}[!ht]
\centering
\caption{AUPRC scores across models and feature sets.}
\setlength{\tabcolsep}{5pt}
\begin{tabular}{|l+l|l|l|l|}
\hline
& \multicolumn{3}{c|}{\textbf{Dataset}} \\ \cline{2-4}
\textbf{Model} & \textbf{Merged} & \textbf{Mobility} & \textbf{Genotype} \\ 
\hline
LOGIT & 0.67 (3.3) & 0.64 (3.0) & 0.60 (3.0)\\
SVC & 0.67 (3.3) & 0.64 (3.0) & 0.60 (3.0)\\
KNN & 0.68 (3.3) & 0.71 (3.4) & 0.62 (3.1)\\
DT & 0.75 (3.6) & 0.76 (3.6) & 0.61 (3.1)\\
RF & 0.91 (4.3) & 0.91 (4.3) & \textbf{0.65 (3.2)}\\
ADA & 0.91 (4.4) & 0.91 (4.3) & 0.61 (3.2)\\
GBC & \textbf{0.93 (4.5)} & \textbf{0.93 (4.4)} & 0.64 (3.2)\\ \hline
\end{tabular}
\begin{flushleft} Mean AUPRC and standard errors ($\times 10^{-2}$) across 100 datasets with co-occurrence $C=0.8$ and relative risks $(G,M)=(\infty,\infty)$.
Bold values indicate models with significantly better AUPRC than other methods for a feature set (Welch's two sample t-test; all p-values $\leq 2.39 \times 10^{-8}$).
\end{flushleft}
\label{tab:prauc}
\end{table}

\begin{table}[!ht]
\centering
\caption{Model $F_1$ scores across models and feature sets.}
\setlength{\tabcolsep}{5pt}
\begin{tabular}{|l+l|l|l|l|}
\hline
& \multicolumn{3}{c|}{\textbf{Dataset}} \\ \cline{2-4}
\textbf{Model} & \textbf{Merged} & \textbf{Mobility} & \textbf{Genotype} \\ 
\hline
LOGIT & 0.63 (2.5) & 0.58 (2.7) & 0.59 (2.9)\\
SVC & 0.63 (2.6) & 0.58 (2.9) & 0.58 (2.9)\\
KNN & 0.52 (4.2) & 0.62 (3.3) & 0.60 (3.4)\\
DT & 0.68 (2.9) & 0.67 (4.3) & 0.59 (3.3)\\
RF & 0.80 (2.3) & 0.80 (2.3) & \textbf{0.66 (3.1)}\\
ADA & 0.82 (2.1) & 0.82 (2.2) & 0.59 (3.0)\\
GBC & \textbf{0.84 (1.9)} & \textbf{0.84 (2.3)} & 0.63 (3.2)\\  \hline
\end{tabular}
\begin{flushleft} Average $F_1$ scores and standard errors ($\times 10^{-2}$) across 100 datasets with co-occurrence $C=0.8$ and relative risks $(G,M)=(\infty,\infty)$.
Bold values indicate models with significantly better $F_1$ scores than other methods for a feature set (Welch's two sample t-test; all p-values $\leq 3.29 \times 10^{-7}$).
\end{flushleft}
\label{tab:f1_inf}
\end{table}

\subsubsection*{Disease co-occurrence}
\begin{figure}[!h]
\centering
\includegraphics[width=1.0\textwidth]{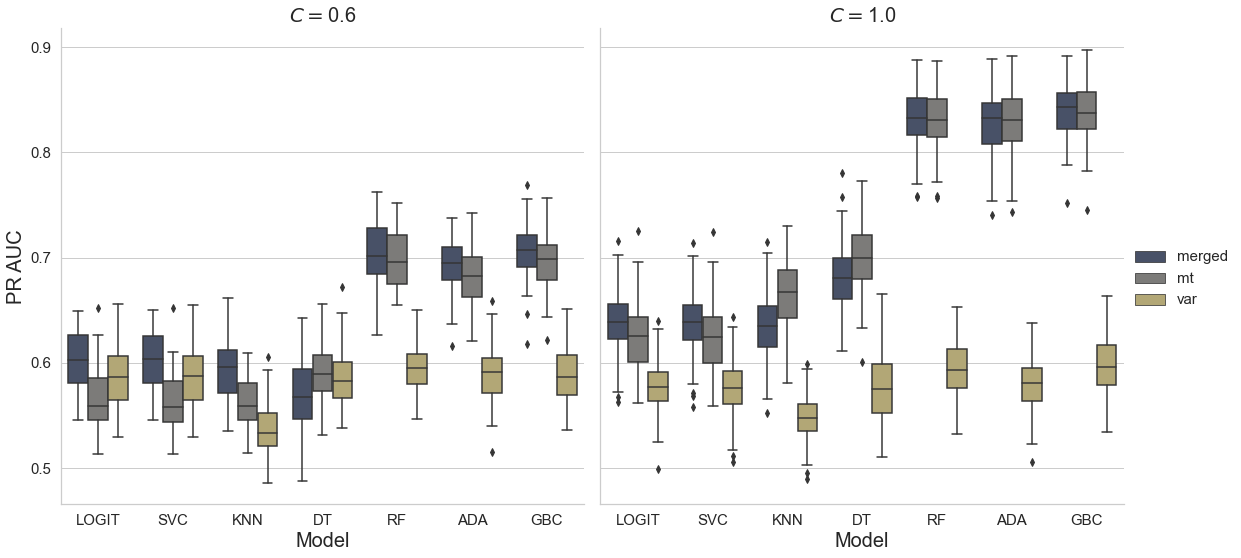}
\caption{{\bf AUPRC across co-occurrence levels.}
Box plots show the distribution AUPRC with Tukey whiskers (median $\pm$ $1.5$ $\times$ interquartile range).
Each method was executed for $100$ datasets across co-occurrence $C\in \{0.6,1.0\}.$ with relative risks $(G,M)=(10,5)$.}
\label{fig:prcom}
\end{figure}

Next, we considered varying the co-occurrence levels between OUD and depression, both to model the uncertainty in co-occurrence between these conditions and to explore how these approaches generalize to different diseases.
The AUPRC across models largely recapitulated the previous AUROC results (Fig.~\ref{fig:prcom}).
Models trained on merged and mobility feature sets had higher AUPRC, though the effect was more pronounced with higher co-occurrence due to the lower noise in case-control status assignment; this more closely models the case where the mobility and genetic features are computed from the same population sample.
Interestingly, the difference between AUPRC in models trained on merged versus mobility trace data is higher in the lower co-occurrence ($C=0.6$), suggesting that the genetic data provides some distinguishability when the case-control assignment is noisier.

\subsubsection*{Relative Risk} 
\begin{figure}[!h]
\centering
\includegraphics[width=1.0\textwidth]{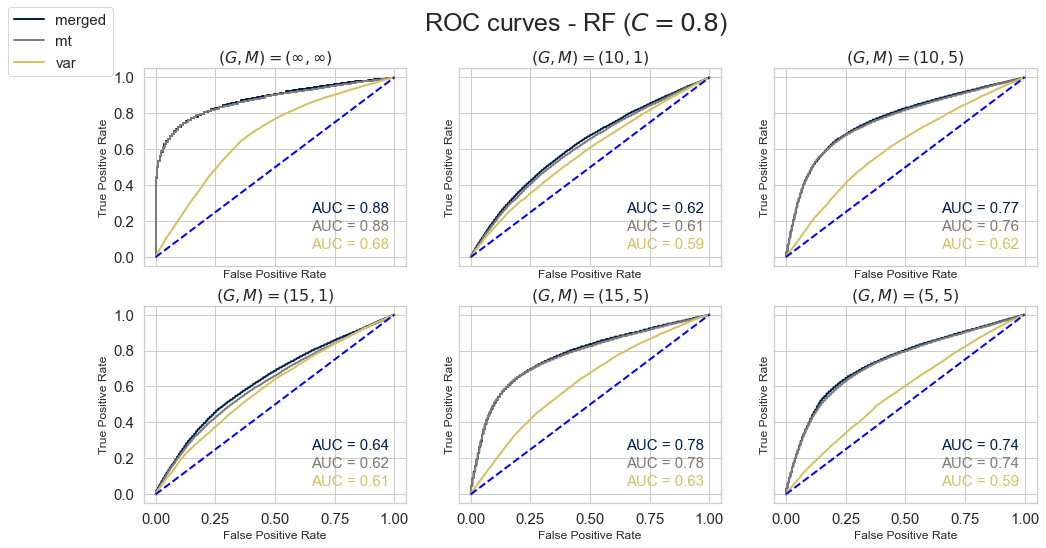}
\caption{{\bf ROC curves across relative risk configurations - random forest.}
Random forest models were trained on $100$ randomly sampled datasets with co-occurrence $C=0.8$ and across relative risks $(G,M) \in \{(\infty, \infty), (10,1), (10,5), (15,1), (15,5), (5,5)\}$.}
\label{fig:rocbyrr}
\end{figure}

\begin{figure}[!h]
\centering
\includegraphics[width=1.0\textwidth]{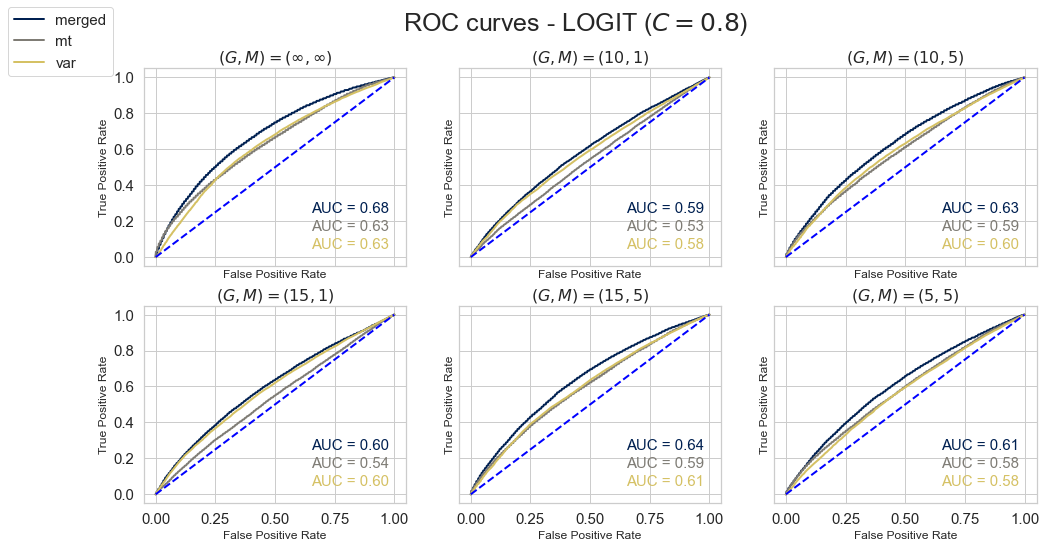}
\caption{{\bf ROC curves across relative risk configurations - logistic regression.}
Logistic regression models were trained on $100$ randomly sampled datasets across relative risks $(G,M) \in \{(\infty, \infty), (10,1), (10,5), (15,1), (15,5), (5,5)\}$.}
\label{fig:rocclog}
\end{figure}

To evaluate the effect of variable disease penetrance from genetic and mobility features, we compared risk model performance across $6$ RR configurations (Tables~\ref{tab:acc_rr} and \ref{tab:f1}).
Consistent with prior results, the mean accuracy and $F_1$ scores show that the ensemble models have the highest performance.
While increasing the RR for both genetic and mobility features increased accuracy, the relative increase was higher for mobility traces.
Next, we averaged the ROC curves across the 100 randomly sampled datasets and RR configurations for a fixed $C=0.8$ and plotted the results, which include RF (Fig.~\ref{fig:rocbyrr}), and logistic regression (Fig.~\ref{fig:rocclog}) classifiers.
The ROC curves also demonstrate that the largest gains in AUROC are achieved with higher mobility RR, but also that genotype RR distinguishes ROC curves for a fixed mobility RR. 
Overall, classification accuracy was diminished when compared with AUROC, where models performed relatively well at ranking case-control probabilities.

\begin{table}[h]
\caption{Model accuracy across relative risks.}
\setlength{\tabcolsep}{4pt}
\begin{tabular}{|l|l|l|l|l|l|l|}
\hline
\textbf{Model} & ($\infty$, $\infty$) & (10,1) & (10,5) & (15, 1) & (15,5) & (5,5)\\
\hline
LOGIT & 0.63 (2.4) & 0.56 (2.8) & 0.59 (2.1) & 0.57 (2.6) & 0.60 (2.5) & 0.58 (2.8) \\
SVC & 0.63 (2.4) & 0.56 (2.8) & 0.59 (2.0) & 0.57 (2.6) & 0.60 (2.2) & 0.58 (2.6) \\
KNN & 0.62 (2.2) & 0.55 (2.5) & 0.59 (2.3) & 0.55 (2.4) & 0.59 (2.3) & 0.57 (2.4)\\
DT & 0.69 (2.9) & 0.53 (2.9) & 0.61 (3.2) & 0.53 (2.4) & 0.61 (2.7) & 0.61 (2.6)\\
RF & 0.82 (2.0) & \textbf{0.59 (2.3)} & \textbf{0.72 (2.3)} & \textbf{0.61 (2.5)} & \textbf{0.73 (2.2)} & \textbf{0.70 (2.3)}\\
ADA & 0.83 (2.0) & \textbf{0.59 (2.8)} & 0.71 (2.3) & \textbf{0.60 (2.6)} & \textbf{0.73 (2.2)} & 0.69 (2.4)\\
GBC & \textbf{0.85 (1.8)} & \textbf{0.59 (2.2)} & \textbf{0.72 (2.4)} & \textbf{0.60 (2.3)} & \textbf{0.73 (2.2)} & \textbf{0.70 (2.3)}\\ 
\hline
\end{tabular}
\begin{flushleft} Mean accuracy and standard errors ($\times 10^{-2}$) across 100 datasets and genotype and mobility RR $(G,M)$, co-occurrence $C=0.8$, and combined genetic and mobility features. 
Bold values indicate models with significantly better accuracy than other non-bolded accuracies within a feature set (Welch's two sample t-test; all p-values $\leq 2.9 \times 10^{-2}$).
\end{flushleft}
\label{tab:acc_rr}
\end{table}

\begin{table}[!ht]
\begin{adjustwidth}{-1.00in}{0in}
\centering
\caption{Model $F_1$ scores across relative risks.}
\setlength{\tabcolsep}{5pt}
\begin{tabular}{|l|l|l|l|l|l|l|}
\hline
\textbf{Model} & ($\infty$, $\infty$) & (10,1) & (10,5) & (15, 1) & (15,5) & (5,5)\\
\hline
LOGIT & 0.63 (2.5) & 0.54 (3.2) & 0.58 (2.4) & 0.56 (3.0) & 0.58 (3.0) & 0.57 (3.1) \\
SVC & 0.63 (2.6) & 0.54 (3.2) & 0.58 (2.4) & 0.55 (3.0) & 0.59 (2.7) & 0.57 (3.0) \\
KNN & 0.52 (4.2) & 0.49 (3.9) & 0.51 (4.3) & 0.50 (3.3) & 0.50 (4.0) & 0.50 (3.6)\\
DT & 0.68 (2.9) & 0.50 (5.8) & 0.57 (4.5) & 0.53 (3.1) & 0.61 (3.3) & 0.54 (6.9)\\
RF & 0.80 (2.3) & 0.58 (2.8) & 0.70 (2.8) & 0.58 (2.8) & 0.71 (2.7) & 0.68 (2.6)\\
ADA & 0.82 (2.1) & 0.57 (3.2) & 0.70 (2.7) & 0.58 (2.9) & 0.71 (2.6) & 0.67 (2.7)\\
GBC & \textbf{0.84 (1.9)} & \textbf{0.59 (2.7)} & \textbf{0.71 (2.5)} & \textbf{0.60 (2.7)} & \textbf{0.72 (2.5)} & \textbf{0.69 (2.4)}\\  \hline
\end{tabular}
\begin{flushleft} Mean $F_1$ scores and standard errors ($\times 10^{-2}$) across 100 datasets with co-occurrence $C=0.8$, genotype and mobility RR $(G,M)$, and combined genetic and mobility features. 
Bold values indicate models with significantly better $F_1$ scores than other methods for a feature set (Welch's two sample t-test; all p-values $\leq 3.1 \times 10^{-3}$).
\end{flushleft}
\label{tab:f1}
\end{adjustwidth}
\end{table}

\subsubsection*{Model interpretation} 
Lastly, we investigated how features shaped model predictions using SHapely Additive exPlanations (SHAP): a model-agnostic and cooperative game-theoretic approach for interpreting fitted models and quantifying feature importance based on feature contribution to a sample prediction~\cite{NIPS2017_7062}. 
The feature values of a data instance $x$ act as players in a coalition, and the prediction output $f(x)$ represents the value generated by the players. 
Shapely values aim to fairly allocate the contribution of each feature $x_j$ to the prediction output.
The SHAP method represents the Shapely value explanation as an additive feature attribution method, a linear model  $g$, which is an interpretable approximation of the original prediction model $f$.
The explanation model is defined as $g(z^{\prime}) = \phi_0 + \sum_{j=1}^{M}\phi_j z^{\prime}_j$ where $z^{\prime} \in \{0,1\}^M$ selects a subset of features, $M$ is the subset size, and $\phi_j \in \mathbb{R}$ is the feature effect attribution for a feature $j$, the Shapely values~\cite{NIPS2017_7062}.

\begin{figure}[!h]
\centering
\includegraphics[width=0.6\textwidth]{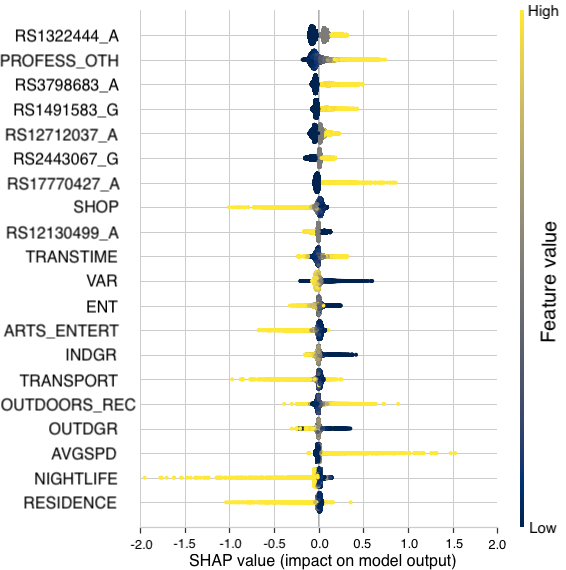}
\caption{{\bf Feature importance vs. effects - SVC.}
SHAP and feature values are shown across the $20$ features with the largest mean difference between expected and predicted values for linear SVC.
Each point represents a SHAP value for a specific feature and instance (colored by the feature value), the y-axis denotes the features ordered according to their average contribution, and the x-axis denotes the SHAP value. 
SHAP values are computed from 100 datasets with co-occurrence $C=0.8$ and relative risks $(G,M)=(10,5)$.
SHAP values that exceeded bounds were excluded.
}
\label{fig:shapdp}
\end{figure}
\begin{figure}[!h]
\centering
\includegraphics[width=0.6\textwidth]{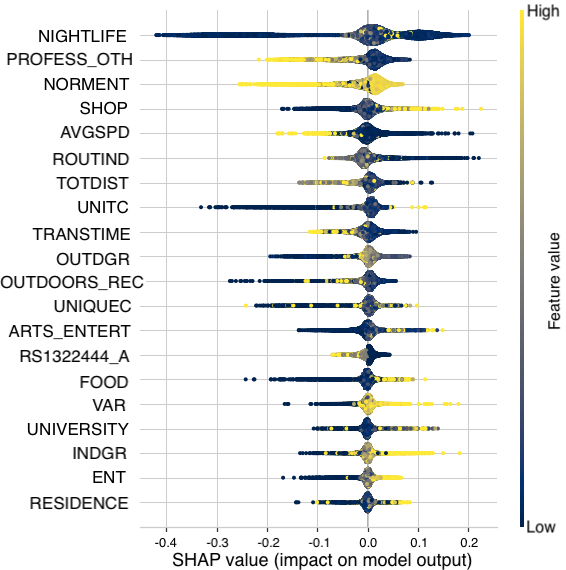}
\caption{{\bf Feature importance vs. effects - random forest.}
SHAP and feature values are shown across the $20$ features with the largest mean difference between expected and predicted values for RF.
Each point represents a SHAP value for a specific feature and instance (colored by the feature value), the y-axis denotes the features ordered according to their average contribution, and the x-axis denotes the SHAP value. 
SHAP values are computed from 100 datasets with co-occurrence $C=0.8$ and relative risks $(G,M)=(10,5)$.
}
\label{fig:shapspSUPP}
\end{figure}

We used a fast implementation of SHAP, TreeSHAP, to explain our tree-based models, and chose the \verb|auto| algorithm parameter for the single model explainers to optimize training time~\cite{NIPS2017_7062}. 
We computed SHAP values for a representative ensemble (RF) and linear (SVC) model; we did not generate explanation results for AdaBoost and KNN since they were not natively supported by \verb|shap v.0.39|.
The output is visualized using SHAP summary plots, which combine feature importance and feature effects. 
We observed that variants had larger mean differences between expected and predicted values (a measure of model impact) for linear SVC compared to RF~(Figs.~\ref{fig:shapdp} and \ref{fig:shapspSUPP}).
In contrast, mobility features had higher SHAP value variability compared to SNPs across both models (a measure of importance).

\begin{figure}[!h]
\centering
\includegraphics[width=1.0\textwidth]{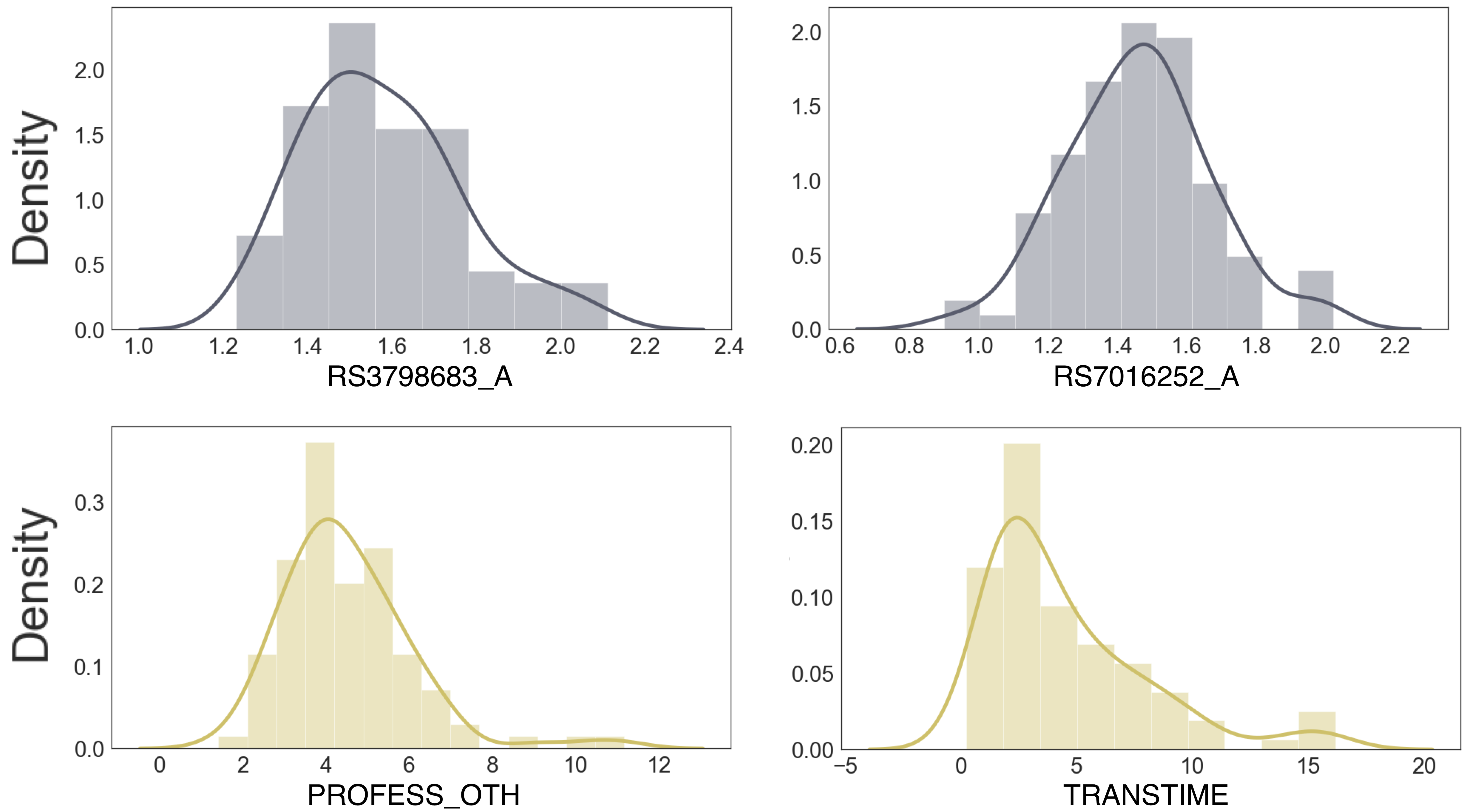}
\caption{{\bf Odds ratios for logistic regression models.}
Odds ratio densities derived from fitted LR models ordered left to right by highest median odds ratio.
Each plot used 100 datasets with combined features, co-occurrence $C=0.8$, and RR $(G,M)=(10,5)$. 
}
\label{fig:odds}
\end{figure}

We also calculated the Pearson correlation coefficients ($\rho$) between SHAP and feature values, which were all significant for RF and SVC (Table~\ref{tab:pearson}; $\rho \neq 0$, p-values $\leq 1.8 \times 10^{-3}$).
Genetic variants were among the highest correlated features (in absolute value) in both RF and SVC models.
Interestingly, while RF models achieved better performance, the linear SVC correlations were more consistent with prior results; for example, entropy of movement (ENT) and amount of time spent home (HOME) have been shown to be negatively and positively correlated with OUD~\cite{farhan_behavior_2016}~(Table~\ref{tab:pearson}).
Location variance and time spent in shop \& service (SHOP) or arts \& entertainment (ARTS\_ENTERT) locations also exhibited significant negative correlations with OUD. 
Lastly, we considered odds ratios computed from fitted logistic regression models, which indicated that exposure to genetic features show a maximum median increase of 55\% in the odds of OUD  (Fig.~\ref{fig:odds}).
In contrast, the median odds of OUD increased approximately fourfold in the two most significant mobility features (time spent in professional \& other places and transition time).

\begin{table}[h!]
\caption{Pearson correlation for SHAP and feature values for random forest and SVC models.}
\setlength{\tabcolsep}{4pt}
\begin{tabular}{|l|l|l|}
\hline
\textbf{Model} & Random Forest & Support Vector Classifier\\ 
\thickhline
VAR & 0.5* & -0.88* \\
AVG SPD & -0.17* & 0.86* \\
ENT & 0.45* & -0.79* \\
NORM ENT & 0.17* & -0.61* \\
HOME & -0.16* & 0.59*\\
TRANS TIME & -0.50* & 0.80*\\
TOT DIST & 0.54* & 0.62*\\
ROUT IND& -0.15* & -0.14*\\
INDGR & 0.29* & -0.75*\\
OUTDGR & 0.24* & 0.59*\\
UNIQUEC & 0.05* & 0.54*\\
UNITC & 0.02* & -0.09*\\
OUTDOORS\_REC & -0.30* & 0.81*\\
PROFESS\_OTH & -0.74* & 0.97*\\
SHOP & 0.39* & -0.92*\\
FOOD & 0.23* & -0.60*\\
TRANSPORT & 0.15* & -0.76*\\
RESIDENCE & 0.09* & -0.82*\\
UNIVERSITY & 0.25* & -0.37*\\
ARTS\_ENTERT & 0.29* & -0.86*\\
NIGHTLIFE & 0.05* & -0.70*\\
RS12130499\_A & 0.60* & -0.88*\\
RS12712037\_A & -0.51* & 0.89*\\
RS1491583\_G & -0.68* & 0.93*\\
RS2443067\_G & -0.61* & 0.91*\\
RS1322444\_A & -0.75* & 0.97*\\
RS3798683\_A & -0.71* & 0.95*\\
RS17770427\_A & -0.72* & 0.94*\\
RS10504659\_A & -0.73* & 0.94*\\
RS16939567\_C & -0.69* & 0.94*\\
RS1554347\_G & -0.70* & 0.94* \\ 
\hline
\end{tabular}
\begin{flushleft} The asterisk (*) denotes a significant positive or negative correlation, which was calculated using a two-sided test with the null hypothesis generated using the exact distribution of the Pearson correlation coefficient (all p-value$\leq 2.2 \times 10^{-16}$).
\end{flushleft}
\label{tab:pearson}
\end{table}

\section*{Discussion}
This work introduced and evaluated a strategy of risk modelling that incorporates both inherent genetic and dynamic mobility features contributing to disease risk.
Our risk models are general purpose and can thus incorporate conventional features informative of OUD risk like SOAPP~\cite{butler_validation_2004,martel_catastrophic_2013}.
We envision that this approach may be a useful component for clinicians to form personalized treatment strategies.
There exists two mechanisms to accommodate this \textit{a priori} use case: (a) if it is known that the patient requires analgesics in advance, mobility data could be generated prior to prescription; and (b) if the injury is acute, the patient could be prescribed alternative analgesics while mobility is being evaluated, though care should be taken to control for situations where the injury affects mobility.
If treatment with opioids cannot be delayed, mobility traces can be used to monitor disease risk over time and inform intervention strategies due to the dynamic nature of disease risk~\cite{giannoula2018identifying}.

There are several opportunities to address key limitations of our work.
First, because mobility trace data does not exist for OUD, a combined genetic and mobility analysis required the fusion of datasets on different samples with related, yet distinct conditions.
Collecting mobility trace and genetic data on the same population sample would address this limitation. 
Additionally, ensemble models achieved high performance with complete penetrance and a co-occurrence of $1$ ($\approx 0.92$ AUPRC).
This is likely an artifact of a small mobility trace sample size and a resulting underestimation of mobility trace population variance.
Thus, increasing the size and cohort diversity of the mobility trace data should improve estimates of generalization performance.
In terms of clinical support, while the risk models generally performed well with respect to AUROC, the ability to rank samples by class probability is less useful in clinical settings where treatment and monitoring decisions are discrete. 
Additionally, recent studies have shown the utility of including clinical biomarkers and other  environmental factors when evaluating substance use risk~\cite{barr2022clinical,kinreich2021predicting}.
Future work should evaluate how these additional modalities interact with features computed from mobility traces.
Finally, while genotypes are clearly causal factors for OUD, the relationship between features extracted from mobility traces and OUD is less clear.
Since depression and OUD are diseases that are commonly presented together and depression status can be imputed from mobility traces with high accuracy, mobility is likely informative of OUD status; however, whether mobility is predictive of OUD risk still needs to be evaluated.

Many concerns must be addressed prior to clinical implementation.
Collecting and storing genomic and mobility data imposes risks to patients, and so disease risk modelling system using mobility traces must address security and privacy concerns. 
Specifically, genomic data informs disease association and forensic identification for both the patient and their relatives~\cite{naveed2015privacy}. 
Spatiotemporal mobility data are highly unique and facilitate the identification of individuals using coarse and sparsely sampled data~\cite{de2013unique}.
Mechanisms to mitigate privacy and security issues must be implemented both at training and inference time~\cite{berger2019emerging,chakraborty2013framework}.
For example, training these models requires a participating cohort to transmit their mobility data to a centralized server. 
However, after a model is trained, patient data can be stored locally (e.g., on their phone) which can also be used to run the trained model and transmit the prediction results to the centralized server. 
After models are trained or mobility features extracted, raw data can be deleted. 
This can be done, e.g., on a weekly basis, removing the need for long term storage of sensitive data.

Clinician decision making is vulnerable to bias, which may lead to suboptimal diagnosis and treatment outcomes~\cite{bornstein2001rationality}.
Furthermore, computational modelling requires training data that may not be representative of the patient population or exhibit historical bias~\cite{mehrabi2021survey}.
To identify bias, a clinical evaluation of risk supported by computational modelling should be \textit{interpretable} by the doctor and \textit{explainable} to both the doctor and patient.
Unfortunately, machine learning techniques generally produce risk assessments that are difficult to explain based on models that are hard to interpret. 
Recent work in machine learning has pursued methods to explain how a machine learning algorithm arrived at a prediction (\textit{explaining} a ``black box'')~\cite{molnar2020interpretable}, or to build machine learning models that humans can understand more easily (an \textit{interpretable} ``glass box''), ideally without sacrificing prediction accuracy~\cite{rudin2021interpretable,lage2018human}.
The advantages and disadvantages of models with respect to accuracy, interpretability, and explainability must be considered when using such a framework in practice.

Explainability techniques have three primary aims: (1) improve user understanding of the model; (2) communicate the uncertainty underlying a model prediction to users (domain experts or laypersons) to lead users to rely on more certain predictions; and (3) help users calibrate their trust in the model appropriately so as to maximize the joint performance of users and models~\cite{bansal2021does}.
Our use of SHAP exemplifies a \textit{model-agnostic feature explanation}~\cite{NIPS2017_7062}, however other methods exist for explaining machine learning predictions that can be evaluated within our framework, including model-specific feature importances~\cite{menze2009comparison,altmann2010permutation}, counterfactual explanations, which are similar to human explanations~\cite{byrne2019counterfactuals,bhatt2020explainable}, and nearest-neighbors methods that identify the most similar data samples~\cite{cover1967nearest}.
Further, we included a diversity of risk score models, in part, to evaluate performance as a function of model interpretability.
Logistic regression provides odds ratios, per feature weight parameters, and uncertainty estimates, but suffered from poor performance.
Conversely, while AdaBoost and GBC do not provide the same aids to interpretation or uncertainty quantification~\cite{malinin2020uncertainty}, they achieved significantly higher performance.

Clinical trials are necessary to evaluate the efficacy of mobility traces in disease risk estimation for OUD and depression~\cite{friedman2015fundamentals}.
Trials incorporating mobility data could also evaluate other disease associated risks, for example, the risk of relapse or response to treatment of those undergoing opioid agonist therapy.
Since OUD is linked to opioid exposure, these trials can focus on a broader range of variants and environmental factors that increase risk of opioid exposure, like impulsivity or antisocial personality disorder~\cite{kock2018personality}.
Additionally, these trials will be necessary to evaluate (a) if the utility of mobility features transfer to other diseases or disorders and (b) the degree to which treatment, mobility, genetic variation, and disease status interact.

\section*{Conclusion}

We presented the first computational approach to disease risk modelling that combines genetic features with mobility traces and applied our methods to opioid use disorder.
Our pipeline extracts $21$ diverse mobility features -- including $9$ newly proposed features -- based on Google Places API, provides algorithms to synthesize new mobility trace samples, and simulates fused genetic and mobility trace samples using customizable co-occurrence and relative risk parameters. 
We demonstrated that (a) combining genetic and mobility features yielded the best performing models across a variety of measures; (b) the ensemble classifiers outperformed more interpretable linear models; (c) the newly proposed features based on Google Places categories had high influence on predictions; and (d) the interactions between mobility and genetic features were consequential.
Also, our approach is highly scalable as DNA sequencing and genotype array costs continue to diminish while cell phones are increasingly ubiquitous. 
While there exists privacy, security, bias, and generalization concerns, we believe that accurately quantifying both the inherent and dynamic components of OUD risk \textit{before and during treatment} could better empower clinicians to use estimated disease risk as a component in a more informed decision making process.

\section*{Acknowledgments}
D.A. and S.L. were supported in part by the University of Connecticut’s Institute for Collaboration on Health, Intervention, and Policy (awarded to D.A.); B.W. was supported in part by the National Science Foundation grant IIS-1407205 (awarded to B.W.).


%
%
%

\end{document}